\documentclass[10pt,twocolumn,letterpaper]{article}
\pdfoutput=1

\usepackage[utf8]{inputenc}
\usepackage{graphicx}
\usepackage{cite}

\usepackage{physics}
\usepackage{upgreek}
\usepackage{color}
\usepackage{amssymb}
\usepackage{gensymb}
\usepackage{floatrow}
\usepackage{authblk}
\usepackage{titlesec}
\renewcommand{\thesection}{\Roman{section}}
\titleformat{\section}{\Large\scshape}{\thesection}{0.7em}{}

\usepackage[
 format=plain, % Die Beschriftung als Absatz.
 labelsep=period,
 labelfont=bf, % Der Bezeichner soll groß und fett geschrieben werden.
 ]{caption}

\begin{document}

\date{\today}% It is always \today, today,

\title{Spin wave propagation in a ring-shaped magnonic waveguide}
\author[1,2]{Franz Vilsmeier\footnote{franz.vilsmeier@univie.ac.at}}

\author[1,3]{Takuya Taniguchi}
\author[1]{Michael Lindner}
\author[1]{Christian Riedel}
\author[1]{Christian Back}
\affil[1]{\textit{Physik-Department, School of Natural Sciences, Technische Universität München, Garching, Germany}}
\affil[2]{\textit{Fakultät für Physik, Universität Wien, Wien, Austria}}
\affil[3]{\textit{Institute of Multidisciplinary Research for Advanced Materials, Tohoku University, Sendai, Japan}}

\date{\today}% It is always \today, today,

\maketitle

\begin{abstract}
\noindent We experimentally investigate frequency-selective spin wave (SW) transmission in a micrometre-scale, ring-shaped magnonic resonator integrated with a linear Yttrium Iron Garnet (YIG) stripe. Using super-Nyquist-sampling magneto-optical Kerr effect microscopy (SNS-MOKE) and micro-focused Brillouin light scattering ($\upmu$-BLS), we probe SW dynamics in the dipolar regime under in-plane magnetisation. Spatially resolved measurements reveal a sharp transmission peak at 3.92\,GHz for an external field of 74\,mT, demonstrating strong frequency selectivity.

Our results show that this selectivity arises from scattering and interference between multiple SW modes within the ring. These modes are governed by the anisotropic dispersion relation, transverse mode quantisation due to geometric confinement, and inhomogeneities of the effective magnetic field. In addition, the anisotropy enforces fixed group velocity directions, leading to caustic-like propagation that limits efficient out-coupling. Fourier analysis reveals discrete wavevector components consistent with quantised transverse eigenmodes. Additional $\upmu$-BLS measurements at 70\,mT show a shift of the transmission peak, confirming that the filtering characteristics are tunable by external parameters.
\end{abstract} %%%%%%%%%

\bigskip

\section{Introduction}

Considerable interest has been drawn to the field of magnonics~\cite{Serga2010,Krawczyk2014,Chumak2015,Barman2021,Chumak2022,Flebus2024}, which aims to study and exploit the properties of spin waves (SWs)—the collective excitations of magnetic moments within a magnetically ordered material. Their potentially nanoscale wavelengths~\cite{Yu2016_2} and tunable dispersion relations make SWs attractive for wave-based information processing, where functionalities such as frequency filtering~\cite{Qin2021,Zenbaa2025} and logic operations~\cite{Lee2008,Goto2019,Papp2021} can be achieved based on wave interference effects.

Self-interference-based SW filters have gained interest due to their structural simplicity and compatibility with standard fabrication processes. A minimal implementation is a ring-shaped magnonic resonator connected to a waveguide, offering frequency-dependent interference paths without multilayer designs, similar to photonic ring resonators~\cite{Rabus2007}. While several theoretical and simulation studies~\cite{Venkat2017,Odintsov2019,Wang2020} and some experimental work~\cite{Iwaba2023,Odintsov2024} have explored such structures, spatially resolved observations of SW propagation and interference in micrometre-scale rings have largely been missing.

In this report, we experimentally investigate SW propagation in a micrometre-scale ring-shaped magnonic resonator connected to a yttrium iron garnet (YIG) waveguide using super-Nyquist-sampling magneto-optical Kerr effect (SNS-MOKE) microscopy. Our spatially resolved measurements reveal that multiple transverse width modes and caustic-like directional SWs coexist and interfere within the ring, giving rise to pronounced frequency-selective transmission. Additional micro-focused Brillouin light scattering ($\upmu$-BLS) experiments were performed, confirming that the ring-structure realized a high frequency selectivity. These findings suggest that the inherent complexity of magnonic mode structure can actively enhance the frequency selectivity of self-interference-based SW filters.

\begin{figure}[h!]
    \includegraphics[width=\textwidth]{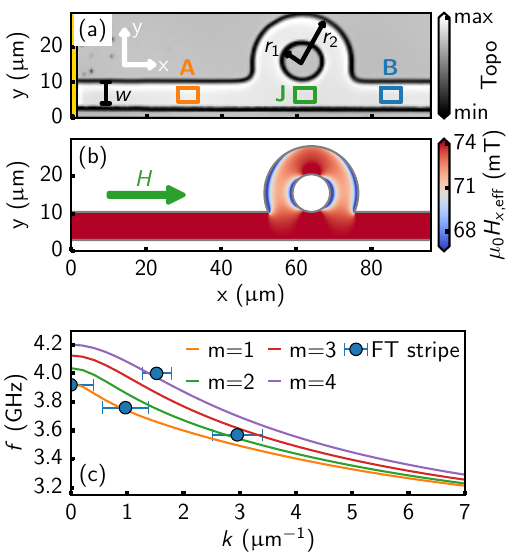}% Here is how to import EPS art
    \caption{\label{fig:Figure1} (a) Topographic image of the ring-shaped magnonic waveguide. Both stripe and ring have a width of $w = 7.5\,\upmu$m; the ring has an inner radius of $r_1 = 5\,\upmu$m and an outer radius of $r_2 = 12.5\,\upmu$m. A 2\,$\upmu$m-wide gold microstrip antenna (gold cuboid) excites SWs in the BV geometry ($\vb*H\perp \vb*e_x$). Rectangles A, J, and B mark regions used to assess the SW amplitude before, at, and after the ring junction. (b) Micromagnetic simulation of the $x$-component of the effective field, $\mu_0 H_{x,\mathrm{eff}}$, at $\mu_0 H = 74$\,mT for longitudinal magnetisation. The green arrow indicates the external field direction. (c) Width-mode-dependent dispersion relation in the BV geometry at $\mu_0 H = 74$\,mT. The mode order is labelled by $m$. Blue circles show experimental values extracted from the stripe region.}
\end{figure}

The magnonic waveguide structure was fabricated from a 200\,nm thick YIG film grown by liquid phase epitaxy on a gadolinium gallium garnet (GGG) substrate. Optical lithography followed by argon ion etching was used to pattern a stripe–ring configuration consisting of a 7.5\,$\upmu$m-wide stripe connected to a ring of the same width and a 5\,$\upmu$m inner radius. A 2\,$\upmu$m-wide gold microstrip antenna was deposited on top of the stripe section to coherently excite SWs in the Backward Volume (BV, $\textbf{k} \parallel \textbf{M}$) geometry. A schematic of the device layout and measurement configuration is shown in Fig.~\ref{fig:Figure1}(a). Additionally, Fig.~\ref{fig:Figure1}(b) shows micromagnetic simulations~\cite{Vansteenkiste2014} of the $x$-component of the effective magnetic field, $\mu_0H_{x,\text{eff}}$ for the longitudinally magnetised geometry with an applied external magnetic field of $\mu_0H=74$\,mT.\footnote{The simulation used a grid of $600\times1920\times4$ and a cell dimension of 50\,nm in all three dimensions.} In the stripe section, the effective magnetic field distribution is homogeneous and matches the applied magnetic field. In contrast, within the ring section, demagnetising effects occur, which lead to local reductions and inhomogeneities of the effective magnetic field. 

Furthermore, Fig.~\ref{fig:Figure1}(c) shows the width-order-dependent dispersion relation of a longitudinally magnetised magnonic waveguide with a width of $w=7.5\,\upmu$m in the BV geometry under an external magnetic field of $\mu_0H=74$\,mT. Due to lateral confinement, the wavevector component $k_y$ across the width is quantised as $k_y=m\pi/w$, where $m$ denotes the mode order. This dispersion relation defines the set of SW modes that can, in principle, be excited in the stripe section. The spectrum spans a broad frequency range, with the ferromagnetic resonance (FMR) mode located around 3.92\,GHz. The analytical expression used for these calculations is provided in the supplementary material.

A discussion of the experiments in the Damon–Eshbach (DE, $\textbf{k} \perp \textbf{M}$) geometry can also be found in the supplementary material.

SW propagation was experimentally probed using super-Nyquist-sampling magneto-optical Kerr effect microscopy (SNS-MOKE)~\cite{Dreyer2018,Qin2021,Dreyer2022} and micro-focused Brillouin light scattering ($\upmu$-BLS)~\cite{Fioretto1993,Sermage1979,BorovikRomanov1982,Birt2013,Djemia2001,Demokritov2001,Demokritov2008,Sebastian2015,Demidov2011,Demidov2015}. SNS-MOKE is a modified version of time-resolved magneto-optical Kerr effect microscopy (TR-MOKE)~\cite{Perzlmaier2008,Farle2012,Au2011,Bauer2014,Stigloher2018,Vilsmeier2024}, which enables spatially resolved detection of both amplitude and phase of the dynamic out-of-plane magnetisation component $\delta m_z$. $\upmu$-BLS provides complimentary access to the SW spectrum by analysing thermally or coherently excited magnons through inelastic light scattering. A more detailed description of the experimental setups can be found in the supplementary material.

\section{Experimental results}

\begin{figure*}[h!]
    \includegraphics[width=\textwidth]{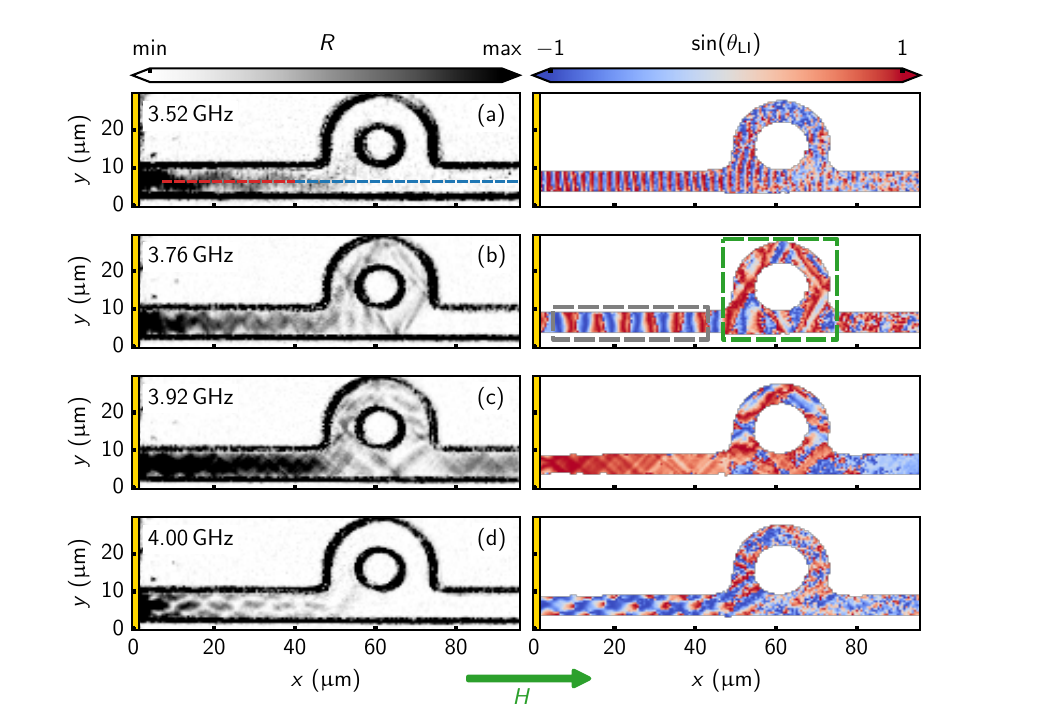}
    \caption{(a)–(d) Kerr maps recorded by SNS-MOKE at different excitation frequencies under an external magnetic field of $\mu_0 H = 74\,$mT. Left column: absolute spin wave amplitude $R = \sqrt{X^2 + Y^2}$ from lock-in detection. Right column: spin wave phase fronts represented as $\sin(\theta_\mathrm{LI})$, where $\theta_\mathrm{LI} = \arctan(Y/X)$. A clear transmission signal beyond the ring is visible in (c). The golden area marks the excitation antenna; the green arrow indicates the external field direction. Line profiles along the red and blue dashed lines in (a) are shown in Fig.~\ref{fig:Figure3}(a). Fourier transform data were extracted from the grey and green contours (stripe and ring) and are shown in Fig.~\ref{fig:Figure1}(c) and Fig.~\ref{fig:Figure3}(b), respectively.}
    \label{fig:Figure2}
\end{figure*}

All the following measurements were carried out in the BV geometry. Figs.~\ref{fig:Figure2}(a)-(d) show spatial Kerr maps recorded by SNS-MOKE at different excitation frequencies, with an external magnetic field of $\mu_0H=74$\,mT applied along the YIG stripe. For each excitation frequency, two spatial maps are provided. The left column displays the absolute precessional amplitude $R=\sqrt{X^2+Y^2}$ where $X$ and $Y$ are the in-phase and out-of-phase components of the lock-in detection signal. These components correspond directly to the real and imaginary parts of the dynamic out-of-plane magnetisation component, $\delta m_z$, thus providing a full description of the SW amplitude. The right column visualises the SW phase in terms of $\sin(\theta_\mathrm{LI})$ where $\theta_\mathrm{LI}=\arctan(Y/X)$. To improve clarity, the phase information is shown only within the device area, as defined by the topographic image, since regions outside exhibit random fluctuations in the absence of SWs.

First, we discuss the propagation profiles of the detected amplitude and phase maps. In Fig.~\ref{fig:Figure2}(a), the SW amplitude decays along the YIG stripe, barely reaching the ring and with no pronounced amplitude beyond it. The phase map shows a short wavelength in the stripe. Several wavefronts are visible inside the ring, and a faint wavefront signature appears beyond the ring.

At an excitation frequency of $f=3.76$\,GHz (Fig.~\ref{fig:Figure2}(b)), plane wavefronts with a longer wavelength propagate along the YIG stripe towards the ring section. Scattering occurs at the transition into the ring, exciting caustic-like features~\cite{Vashkovskii1988,Wartelle2023,Demidov2009,Schneider2010,Riedel2023} that reflect back and forth along nearly the entire ring. A sharp amplitude drop appears at the second ring junction ($x\approx75\,\upmu$m). The phase map reveals complex wavefronts within the ring. Beyond the ring, amplitude is largely absent, though the phase map may indicate a weak transmitted wave.

At $f=3.92$\,GHz (Fig.~\ref{fig:Figure2}(c)), amplitude is observed up to and inside the ring, with a long-wavelength phase pattern. Caustic-like features appear again, now with different beam directions and reflection angles extending into the stripe. Reduced amplitude is observed in the intermediate regions of the ring and stripe. In contrast to the previous cases, a clear amplitude signal is visible beyond the ring, suggesting that transmission along the waveguide depends on the excitation frequency.

At $f = 4.00$\,GHz (Fig.~\ref{fig:Figure2}(d)), a nodal structure across the width of the YIG stripe is observed, pointing to the excitation of higher-order width modes~\cite{Demidov2015,Brcher2017_2,Chumak2019,Taniguchi2024}. No transmission beyond the ring is visible.

To assess the frequency-dependent transmission beyond the ring observed in Figs.~\ref{fig:Figure2}(a)–(d), we evaluated whether the suppression of propagation can be explained solely by the intrinsic attenuation length of the SWs. For this analysis, line profiles of the precession amplitude $R$ were extracted by averaging over three lines along the centre of the stripe. The region before the ring was fitted with an exponential decay of the form
\begin{equation*} 
f(y) = A_\mathrm{exp}\cdot\mathrm{e}^{-x/L_\mathrm{att}}. 
\end{equation*}
Here, $A_\mathrm{exp}$ denotes the amplitude, and $L_\mathrm{att}$ the attenuation length. The dashed lines in Fig.~\ref{fig:Figure2}(a) illustrate the extraction process, where the red and blue dashed lines indicate the regions before and beyond the ring, respectively.

\begin{figure*}[h]
    \includegraphics[width=\textwidth]{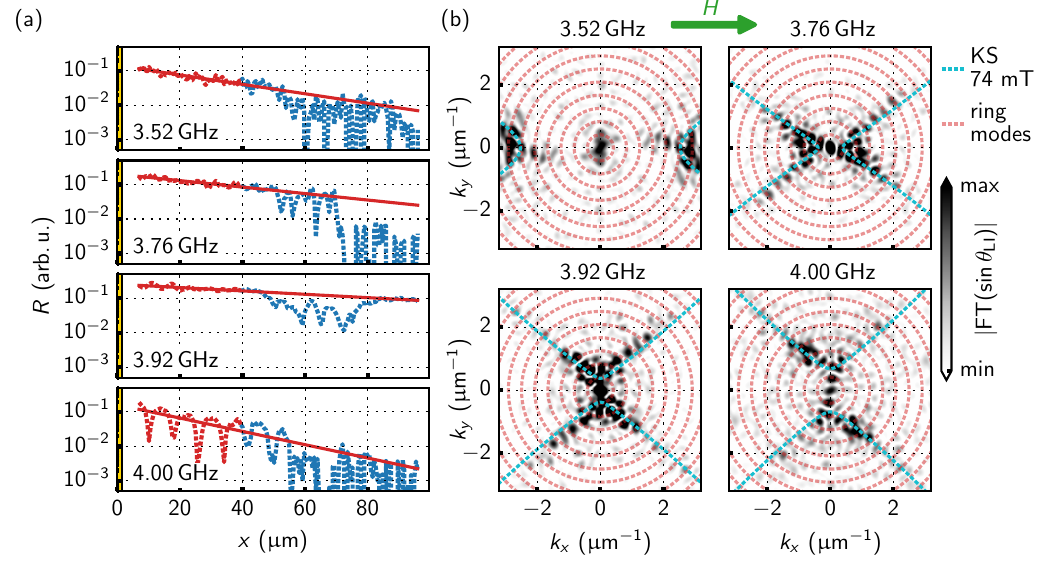}% Here is how to import EPS art
    \caption{(a) Line profiles of the amplitude $R$ for different excitation frequencies, extracted along the dashed red and blue lines in Figs.~\ref{fig:Figure2}(a)–(d). The solid red curve shows an exponential fit to the red profile before the ring section. (b) Fourier transformed (FT) data extracted from the ring region. Red dashed circles indicate the quantised eigenmodes across the ring width. The observed spin wave modes are discretised due to geometric confinement. Cyan dashed lines correspond to the analytical Kalinikos and Slavin (KS) model for $\mu_0H=74\,$mT.}
    \label{fig:Figure3}
\end{figure*}

The resulting profiles are shown in Fig.~\ref{fig:Figure3}(a), where the amplitude $R$ is plotted on a logarithmic scale. The red dotted line corresponds to the extracted profile from the stripe section before the ring, and the blue dotted line shows the profile beyond the ring. The solid red line represents the exponential fit to the pre-ring region.

At excitation frequencies, $f = 3.52$\,GHz and $f = 4.00$\,GHz (top and bottom panels), the absence of amplitude beyond the ring is consistent with the expected exponential decay. At $f = 3.76$\,GHz (second panel), the behaviour deviates. Within the junction region where the stripe and ring merge, the amplitude approximately follows the exponential trend, but local variations are present, which suggest interference of SWs. Beyond the ring, the amplitude drops sharply, deviating significantly from the exponential decay. This indicates that damping alone cannot explain the lack of transmission at this frequency. Moreover, since the suppression occurs at the second junction, it cannot be attributed solely to scattering into the ring arm either, as this process should be symmetric at both junctions. At $f = 3.92$\,GHz (third panel), a clear amplitude signal is detected beyond the ring, consistent with the exponential decay fitted in the stripe, suggesting that magnon-magnon interaction barely affects the transmitted SW. Within the junction, the signal is partially reduced. Taken together, these observations indicate that transmission is not solely determined by intrinsic damping, but also by interference and scattering between SW modes within the ring and at the stripe–ring junctions.

To further investigate the influence of the ring structure on SW propagation and interference, we performed a spatial Fourier transform (FT) of the wavefront maps, separately analysing the stripe and ring sections. The FT data essentially yields a representation of the experimental iso-frequency curves in this region. For each excitation frequency, FT data were extracted from the stripe section within the grey dashed contour and from the ring section within the green dashed contour in Fig.~\ref{fig:Figure2}(b).

In the stripe section, predominantly a single SW mode is observed. The corresponding extracted wavenumbers are plotted in Fig.~\ref{fig:Figure1}(c). The FT data for 3.52\,GHz, 3.76\.GHz and 3.92\,GHz align well with the $m=1$ width mode, while the data at 4.00\,GHz match the $m=4$ width mode, confirming the excitation of higher-order transverse modes~\cite{Demidov2015,Brcher2017_2,Chumak2019,Taniguchi2024}.

In contrast, the FT spectra in the ring section, shown in Fig.~\ref{fig:Figure3}(b), reveal multiple SW modes. The FT spectra are compared to analytically calculated iso-frequency contours based on the Kalinikos–Slavin (KS) model for SWs in soft ferromagnetic thin films~\cite{Kalinikos1986}. The calculations were performed for an external magnetic field of $\mu_0 H = 74$\,mT and are shown as dotted cyan lines. The experimental data show good agreement with the model, indicating that the incident stripe mode scatters into several modes within the ring, governed by the anisotropic in-plane dispersion relation.

The presence of these multiple modes can contribute to the suppression of transmission through two mechanisms. First, interference between the scattered modes within the confined geometry of the ring can lead to destructive interference and local amplitude suppression. Second, the directionality of SW propagation further hampers transmission. Due to stationary group velocity directions imposed by the anisotropic dispersion, SWs travel along fixed angles, forming caustic-like trajectories that undergo multiple internal reflections without efficient out-coupling~\cite{Vashkovskii1988,Wartelle2023,Demidov2009,Schneider2010,Riedel2023}. Additionally, demagnetising effects in the ring lead to local reductions and spatial inhomogeneities in the effective magnetic field (cf. Fig.~\ref{fig:Figure1}(b)), causing further variation in the local dispersion relation and adding complexity to SW behaviour in this region.

Furthermore, the excited wavenumbers in the ring are discrete, as seen most clearly at $f = 3.76$\,GHz and $f = 3.92$\,GHz, where strong amplitude signals are present in Fig.~\ref{fig:Figure3}(b). This discretisation aligns well with the transverse width eigenmodes of the ring. Red dotted circles in Fig.\ref{fig:Figure3}(b) indicate these eigenmodes, calculated as $k_{\mathrm{ring}, m} = m\pi / w_{\mathrm{ring}}$, with $w_{\mathrm{ring}} = 7.5\,\upmu$m and $m = 0, 1, 2, \dots$. These circles coincide with the discrete FT features, indicating that the finite ring width imposes additional confinement. Consequently, the SWs permitted within the ring are determined by the convolution of the dispersion relation with the ring's width eigenmodes.

In summary, the frequency filtering observed in this system does not arise from simple phase accumulation of a single wave mode. Instead, multiple SW modes are excited within the ring, leading to frequency-dependent interference and directionally trapped propagation. The resulting transmission is shaped by the anisotropic dispersion relation, geometric confinement of the propagation spectrum (cf. Fig.~\ref{fig:Figure3}(b)), and local field inhomogeneities (cf. Fig.~\ref{fig:Figure1}(b)), all of which impose strong mode selection within the ring.

\begin{figure*}[t]
    \includegraphics[width=\textwidth]{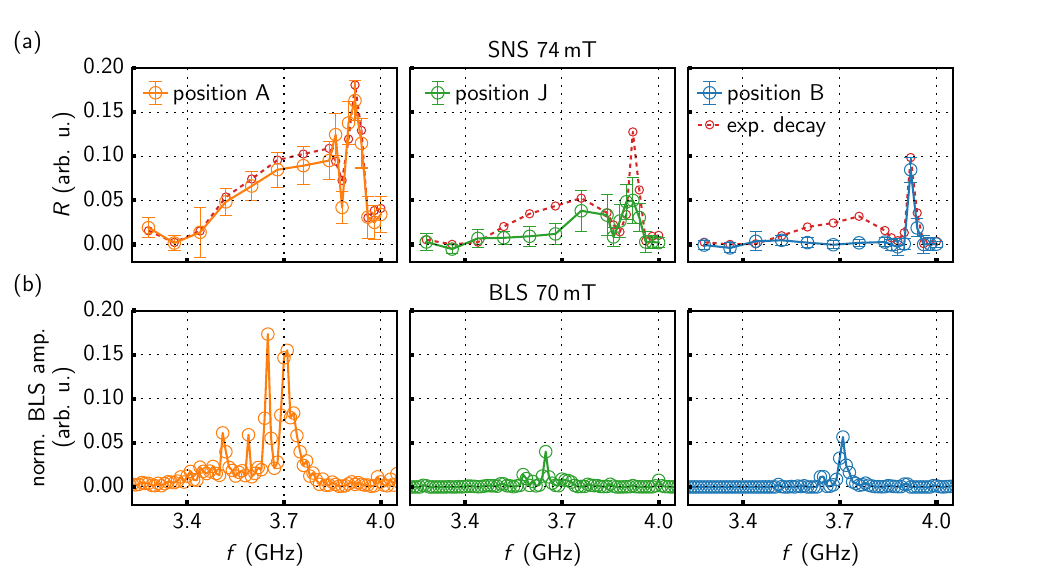}% Here is how to import EPS art
    \caption{(a) SNS amplitudes at $\mu_0 H = 74$\,mT, averaged over regions A, J, and B (cf. Fig.~\ref{fig:Figure1}(a)) as a function of the excitation frequency. Error bars represent the standard deviation of the mean. Small red dots and dashed red lines indicate the expected values at the centre of each region, assuming exponential decay in the stripe before the ring (cf. red fits in Fig.~\ref{fig:Figure3}(a)). (b) $\upmu$-BLS results showing the normalised amplitude of the anti-Stokes peak of the coherent SW excitations. A frequency shift of the transmission peak is observed compared to the 74\,mT case.}
    \label{fig:Figure4}
\end{figure*}

Further SNS-MOKE measurements at different excitation frequencies were performed to assess the transmission properties of the ring. In this context, regions A, J, and B were defined, as illustrated in the topography image in Fig.\ref{fig:Figure1}(a). Averaging the amplitude $R$ over these rectangular regions provides a measure of the experimental SW amplitude before, at, and after the ring junction. The results for various frequencies are presented in Fig.\ref{fig:Figure4}(a). The small red dots indicate the extrapolated values at the centres of regions A, J, and B, obtained from the exponential fit to the stripe section before the ring (cf. Fig.~\ref{fig:Figure3}(a)).

The SW amplitude varies with excitation frequency across different device regions. In region A, the amplitude is low for excitation frequencies below $f \approx 3.5$\,GHz. As the frequency increases, the amplitude gradually rises. A sharp dip occurs at $f = 3.88$\,GHz, followed by a strong increase to a maximum at $f = 3.92$\,GHz. Beyond this point, the amplitude rapidly decreases. The experimental values closely follow the extrapolated amplitudes from the exponential fit.

In region J, the amplitude also increases with frequency, showing a pronounced dip near $f = 3.88$\,GHz and peaking at $f = 3.92$\,GHz. The relatively large error bars in this region indicate complex interference patterns at the junction. The extrapolated amplitudes show generally good agreement with the experimental data, though they tend to slightly overestimate the measured values, particularly at $f = 3.92$\,GHz.

Region B exhibits different behaviour compared to regions A and J. Beyond the ring, little to no amplitude is detected for frequencies below $f = 3.92$\,GHz. At $f=3.92$\,GHz, a sharp transmission peak is observed, rapidly falling off afterwards, exhibiting some amplitude at $f=3.94$\,GHz, but no amplitude beyond. The extrapolated amplitudes from the exponential fit predict sizable values between $f = 3.60$\,GHz and $f = 3.84$\,GHz. However, this prediction contrasts with the experimental results, which show no detectable amplitude in this frequency range. 

Overall, this highlights the system’s sharp frequency selectivity, with a clear transmission peak at $f = 3.92$\,GHz, while other frequency bands exhibit strong attenuation beyond the ring. As previously discussed, the suppression of transmission likely results from interference and scattering among multiple SW modes, similar to the simulations by Odintsov~\textit{et al.}~\cite{Odintsov2019}, rather than from simple phase accumulation along the ring.

Since the interfering modes depend on both frequency and magnetic field, the transmission properties should be tunable by adjusting external parameters. To test this, additional $\upmu$-BLS measurements were performed at an external magnetic field of $\mu_0H = 70$\,mT. The $\upmu$-BLS spectra were acquired at the centres of regions A, J, and B (cf. Fig.~\ref{fig:Figure1}(a)). The results are presented in Fig.~\ref{fig:Figure4}(b), showing the normalised amplitude of the anti-Stokes peak of the coherent SW excitations. Further details on the $\upmu$-BLS measurements and data analysis are provided in the supplementary material.

Similar to the SNS data, position A shows a relatively broad SW spectrum, although it appears less continuous than in previous measurements. At position J, a sharp but weak peak is observed at $f \approx 3.68$\,GHz. More importantly, a clear transmission peak is detected at position B, now shifted to $f = 3.70$\,GHz compared to the 3.92\,GHz peak observed at $\mu_0 H = 74$\,mT. This frequency shift demonstrates that the transmission characteristics can be tuned by varying the external magnetic field.

%%%%%%%%%%%%%%%%%%%%%%%%%%%%%%%%%%%%%%%%%%%%%%%%%%%%%%%%%%%%%%%%%%%%%%%%%%%%%%%%%%%%%%%%%%%%%%%%%%%%%%%%%%%%%%%%%%

\section{Conclusion}

We have experimentally investigated SW transmission in a micrometre-scale, ring-shaped magnonic resonator integrated with a linear YIG stripe. SNS-MOKE and $\upmu$-BLS spectroscopy was used to study SW behaviour in the BV geometry. In the stripe section, primarily a single SW mode was observed, whereas in the ring section, multiple modes emerged and were quantised due to the finite width of the waveguide. Demagnetising effects near and within the ring further modified the local effective magnetic field and influenced SW propagation.

Our results demonstrate that the observed frequency selectivity cannot be explained by a simple exponential decay due to damping or an accumulated phase of a single mode. Instead, transmission is governed by interference between multiple SW modes, shaped by the anisotropic dispersion relation, geometric confinement, and local variations in the effective magnetic field within the ring. In addition, the anisotropy imposes fixed group velocity directions, giving rise to caustic-like propagation that further limits efficient out-coupling.

Spatially resolved SNS-MOKE measurements reveal a pronounced transmission peak at 3.92\,GHz under an applied field of 74\,mT. Additional $\upmu$-BLS measurements confirm that the position of the transmission peak can be tuned by varying the external magnetic field, demonstrating the feasibility of field-controlled frequency selectivity.

%\newpage
%\printbibliography
\bibliography{bibliography}
\bibliographystyle{ieeetr}

\end{document}

\renewcommand{\thefigure}{S\arabic{figure}}

\title{Supplementary material: Spin wave propagation in a ring-shaped magnonic waveguide}

\author[1,2]{Franz Vilsmeier\footnote{franz.vilsmeier@univie.ac.at}}

\author[1,3]{Takuya Taniguchi}
\author[1]{Michael Lindner}
\author[1]{Christian Riedel}
\author[1]{Christian Back}
\affil[1]{\textit{Fakultät für Physik, Technische Universität München, Garching, Germany}}
\affil[2]{\textit{Fakultät für Physik, Universität Wien, Wien, Austria}}
\affil[3]{\textit{Institute of Multidisciplinary Research for Advanced Materials, Tohoku University Sendai, Japan}}

\date{\today}% It is always \today, today,

\maketitle

\section{Experimental method: Super-Nyquist-sampling magneto-optical Kerr effect microscopy}

As a source of illumination, a mode-locked Ti:Sa laser with a centre wavelength of 800\,nm and a pulse width of approximately 150\,fs is used. The pulse trains are applied at a fixed repetition rate of 80\,MHz. The laser polarisation is fixed and the laser focused onto the sample through an objective lens with a numerical aperture of 0.7, giving a spatial resolution of approximately 0.6\,$\upmu$m. Upon reflection at the magnetic surface, the polarization changes due to the polar magneto-optical Kerr effect. This rotation is directly proportional to the dynamic out-of-plane magnetisation component. A Wollaston prism splits the reflected light into orthogonal polarization components, which are detected by two photodiodes. The difference signal corresponds to the Kerr response, while the sum is proportional to the sample’s reflectivity. The sample is mounted on a piezo stage, allowing spatial scanning of the laser focus to generate Kerr maps along with the topography.

We employ super-Nyquist sampling MOKE (SNS-MOKE) to probe SW dynamics, which exploits aliasing in undersampled signals. The laser and rf excitation are phase-locked via a shared reference clock to ensure a constant phase relation. The resulting signal is detected using a \textit{Zurich Instruments HF2LI} lock-in amplifier. By detuning the excitation frequency slightly from an integer multiple of the laser repetition rate, the SW signal is coherently down-converted to a low-frequency alias. This enables lock-in detection without the need for active modulation. Both in-phase and out-of-phase components are acquired in a single measurement, providing access to amplitude and phase information across a wide range of excitation frequencies~\cite{Dreyer2018,Qin2021,Dreyer2022}.

\section{Experimental method: micro-focused Brillouin light scattering}

For the $\upmu$-BLS measurements, a 532\,nm continuous-wave laser is focused onto the magnetic sample through an objective lens with a numerical aperture of 0.75. The beam is incident perpendicular to the sample surface. Inelastic scattering of photons by magnons leads to frequency shifts in the reflected light. The photon loses energy when a magnon is created during the scattering process, which referred to as the Stokes process, resulting in a negative frequency shift of the backscattered light. Conversely, when a magnon is annihilated, the photon gains energy in the Anti-Stokes process, leading to a positive frequency shift~\cite{Fioretto1993,Sermage1979,BorovikRomanov1982,Birt2013,Djemia2001,Demokritov2001,Demokritov2008,Sebastian2015,Demidov2011,Demidov2015}. The frequency shift is analysed by a \textit{TFP-2 HC} tandem Fabry-Perot interferometer~\cite{Scarponi2017} which allows for high contrast over a broad frequency range. This technique allows detection of both thermally and coherently excited magnons; however, in our experiments, we focus exclusively on the coherently driven magnons excited by the rf field. Furthermore, only the Anti-Stokes peak in the frequency shift spectrum is considered in the experiments.

\begin{figure}[h]
    \includegraphics[width=0.7\textwidth]{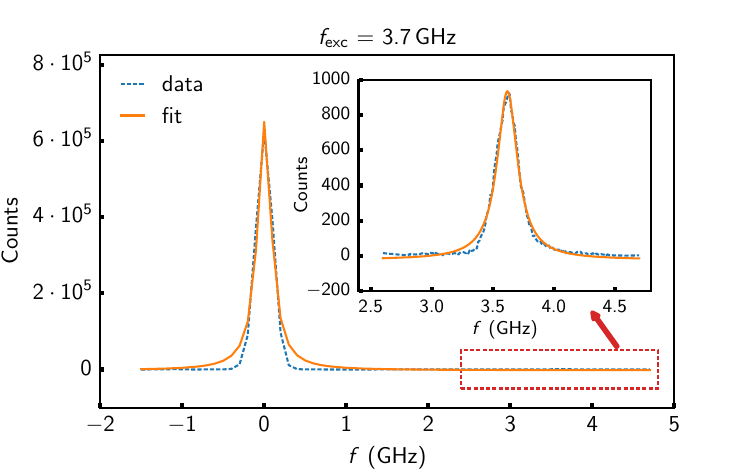} 
    \caption{Detected $\upmu$-BLS spectra. The elastic peak is prominent. The inset displays the Anti-Stokes peak of coherently excited magnons at an excitation frequency of $f_\mathrm{exc}$ and under an external magnetic field of $\mu_0H = 70\,$mT. The blue dotted curve illustrates the experimental data, and the orange curve the Lorentzian fits.}
    \label{fig:SupplementaryMaterial_Figure1}
\end{figure}

Fig.~\ref{fig:SupplementaryMaterial_Figure1} shows a recorded $\upmu$-BLS spectrum of coherently excited magnons at an excitation frequency $f_\mathrm{exc}=3.7$\,GHz and an external field of $\mu_0 H = 70\,$mT. The elastic (zero-shift) peak dominates the spectrum, while the inelastic peak is shown in the inset. For the analysis presented in the main part of the manuscript, both peaks are individually fitted using a Lorentzian function of the form
\begin{equation*}
    L(f) = \frac{A_L \cdot \left( \frac{\Delta f}{2} \right)}{(f - f_0)^2 + \left( \frac{\Delta f}{2} \right)^2} + C,
\end{equation*}
where $A_L$ denotes the amplitude, $f_0$ the resonance frequency, $\Delta f$ the full width at half maximum (FWHM) and $C$ a constant background offset. For the inelastic peak, the amplitude $A_L$ is proportional to the SW intensity. The $\upmu$-BLS data shown in the main part are normalized to the elastic peak amplitude, i.e., $A_{L,\mathrm{inel.}}/A_{L,\mathrm{el.}}$.

\section{Width-dependent dispersion relation in magnonic waveguides}

Following the analytical model by Kalinikos and Slavin~\cite{Kalinikos1986}, the full film dispersion relation in the presence of an in-plane external field  with totally unpinned surface states is given by
\begin{equation}
    \label{eq:dispersion}
    \omega_n^2= \left(\omega_\mathrm{H} + l_\mathrm{ex}^2k_n^2\omega_\mathrm{M}\right)\left(\omega_\mathrm{H} + l_\mathrm{ex}^2k_n^2\omega_\mathrm{M} + \omega_\mathrm{M}F_{n}\right), 
\end{equation}
where
\begin{equation}
    \omega_\mathrm{H}=\gamma\mu_0 H,
\end{equation}
\begin{equation}
    \omega_\mathrm{M}=\gamma\mu_0 M_\mathrm{S},
\end{equation}
\begin{equation}
    F_{n}=P_{n} + \left(1-P_{n}\left(1+\cos^2\varphi\right)+\omega_\mathrm{M}\frac{P_{n}\left(1-P_{n}\right)\sin^2\varphi}{\omega_\mathrm{H}+l_\mathrm{ex}^2k_n^2\omega_\mathrm{M}}\right),
\end{equation}
and
\begin{equation}
\begin{aligned}
    P_{n}= & \frac{k^2}{k_n^2}-\frac{1}{2}\frac{k^4}{k_n^4}F_n, \\
    F_n= & \frac{2}{kL}\left(1-\mathrm{e}^{-kL}\right).
\end{aligned}
\end{equation}
Furthermore, $L$ denotes the film thickness, $k_n=\sqrt{k^2+\left(\frac{n\pi}{L}\right)^2}$, and $\varphi$ describes the angle between $\textbf{k}$ and $\textbf{M}$ (so for $\textbf{k}\parallel\textbf{M}$, $\varphi=0$).

Considering a SW waveguide of finite width $w$, an additional quantisation across the waveguide width is introduced and the dispersion relation can be represented using equ.~(\ref{eq:dispersion}) by letting $k\to\sqrt{k^2+\left(\frac{m\pi}{w}\right)^2}$ and $\varphi\to\varphi-\arctan\left(\frac{m\pi}{kw}\right)$~\cite{Demidov2015,Brcher2017,Chumak2019,Vilsmeier2024}. Here, $m=0,1,2,...$ denotes the eigenmode orders across the width, and $k$ denotes the wavenumber along the waveguide. 

This expression was used to analytically calculate the dispersion relation shown in Fig.~1(a) of the main text.

\section{Verification of discrete mode origin in the ring section}

\begin{figure}[h]
    \includegraphics[width=0.7\textwidth]{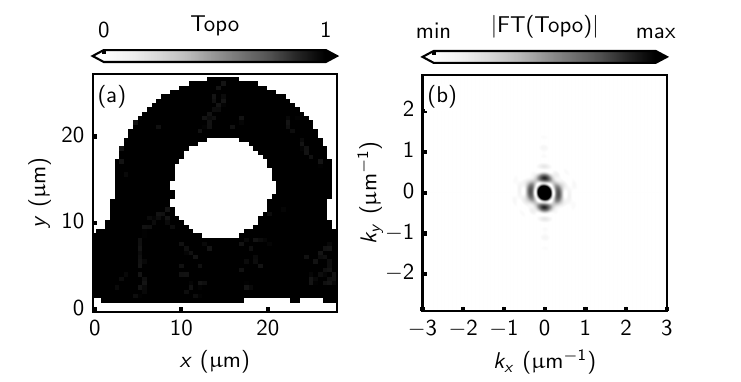} 
    \caption{(a) Binary representation of the ring section topography. (b) Corresponding FT data of the binary image.}
    \label{fig:SupplementaryMaterial_Figure2}
\end{figure}

In the main text, the spatial FT analysis revealed that the allowed wavenumbers within the ring appear to be discrete (cf. Fig.~3(b)). To verify that this discretization arises from the magnetic contrast and not merely from the ring geometry, an additional Fourier transform was performed on a binary image of the ring section. In this image, all areas exhibiting wavefront contrast were assigned a value of 1, while areas without contrast were set to 0, effectively representing only the geometric topography of the ring. This binary mask was then subjected to the same FT procedure used in the analysis of the experimental SW data. Fig.\ref{fig:SupplementaryMaterial_Figure2}(a) shows the binary image, and Fig.\ref{fig:SupplementaryMaterial_Figure2}(b) shows the corresponding FT data. In this case, the FT amplitude is strongly localised around $k = 0,\upmu\text{m}^{-1}$, indicating that no discrete spatial frequencies are present due to the ring geometry alone.

This result supports the conclusion that the discrete wavenumber patterns observed in Fig.~3(b) originate from quantised SW modes and are not artefacts of the sample geometry or image contrast.

\section{Experimental results: Damon-Eshbach geometry}

\begin{figure}[h]
    \includegraphics[width=0.7\textwidth]{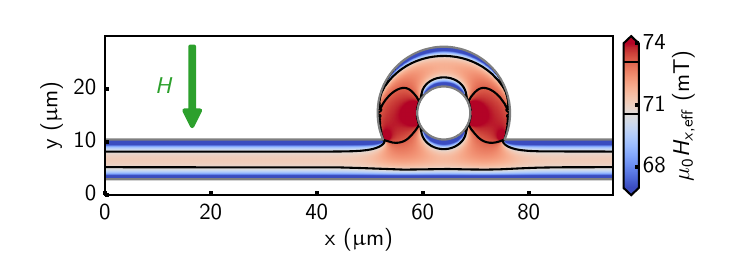} 
    \caption{Micromagnetic simulations of the $y$-component of the effective magnetic field, $\mu_0H_{y,\text{eff}}$, for a transversely magnetised waveguide under an external magnetic field of $\mu_0H=74\,$mT. The black contour lines correspond to effective magnetic fields of $\mu_0H_{y\mathrm{,eff}}=70.6$\,mT and $\mu_0H_{y\mathrm{,eff}}=73.1$\,mT. The green arrow indicates the direction of the applied external magnetic field.}
    \label{fig:SupplementaryMaterial_Figure3}
\end{figure}

Analogous measurements were conducted in the Damon–Eshbach (DE, $\textbf{k} \perp \textbf{M}$) geometry. i.e. with the YIG stripe transversely magnetised. Fig.~\ref{fig:SupplementaryMaterial_Figure3} shows micromagnetic simulations of the effective magnetic field\cite{Vansteenkiste2014}.\footnote{The simulation was performed using a grid of $600\times1920\times4$ cells with a spatial resolution of 50\,nm in all three dimensions.} Again, an external magnetic field of $\mu_0H = 74$\,mT was applied. Compared to the longitudinally magnetised case, the stripe section shows a pronounced reduction in the effective field. As the stripe merges into the ring, the effective magnetic field gradually increases, as indicated by the iso-field line at $\mu_0H_{y,\mathrm{eff}} = 70.6$\,mT. Within the ring, the effective magnetic field remains spatially inhomogeneous, resulting in pronounced local variations in the dispersion relation.

\begin{figure}[h!]
    \includegraphics[width=\textwidth]{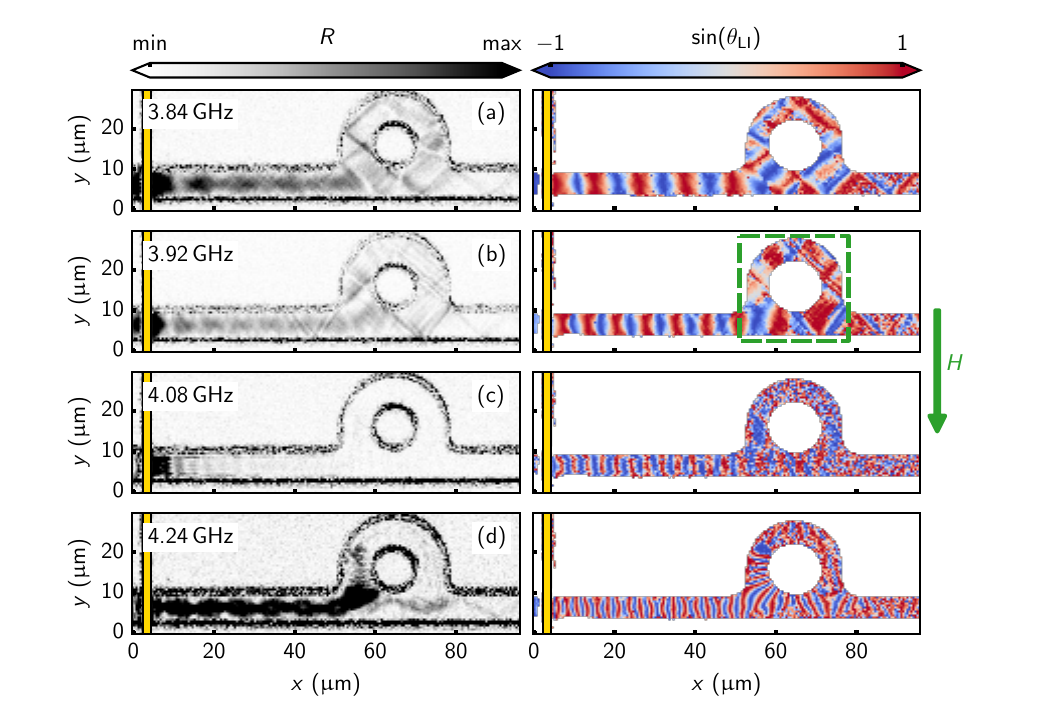}
    \caption{Maps in the DE-geometry at different frequencies with an external magnetic field of $\mu_0H=74\,$mT. (a)-(d) propagation within the stripe is observed where the wavelength decreases with increasing external magnetic field. Caustic-like features are visible in the stripe section after the ring in (a) and (b). FT data extracted from areas within the green contour in (b) is shown in Fig.~\ref{fig:SupplementaryMaterial_Figure6}(b).}
    \label{fig:SupplementaryMaterial_Figure4}
\end{figure}

Experimental data acquired using SNS-MOKE are depicted in Figs.~\ref{fig:SupplementaryMaterial_Figure4}(a)-(d). In each map, a detectable amplitude signal is present within the stripe section. The phase maps confirm that plane SWs are launched from the microstrip antenna at all excitation frequencies. Consistent with the DE dispersion relation, the wavelength decreases with increasing frequency. In Figs.~\ref{fig:SupplementaryMaterial_Figure4}(a) and (b), significant amplitude is observed in both the initial stripe section and the ring. Beyond the ring, the amplitude in the stripe drops sharply, although caustic-like beam features reflecting back and forth remain visible, highlighting the role of direction-dependent scattering in mediating transmission.. As the SWs approach the ring, the phase maps show an increase in wavelength, which can be attributed to the gradual rise in the effective magnetic field near the stripe–ring transition, as seen in the simulations.

In Fig.~\ref{fig:SupplementaryMaterial_Figure4}(c), the amplitude signal vanishes early along the stripe, whereas in Fig.~\ref{fig:SupplementaryMaterial_Figure4}(d), it abruptly decreases after entering the ring. Beam-like features beyond the ring are also observed in Fig.~\ref{fig:SupplementaryMaterial_Figure4}(d).

\begin{figure}[h!]
    \includegraphics[width=0.4\textwidth]{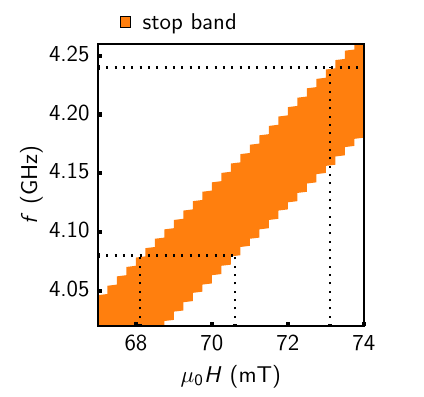}
    \caption{Hybridisation-induced stop band for Damon Eshbach modes. The dotted lines illustrate the respective field limits of the stop band at $f=4.08$\,GHz and $f=4.24$\,GHz.}
    \label{fig:SupplementaryMaterial_Figure5}
\end{figure}

In the DE geometry, an additional factor must be considered to fully describe SW propagation. DE modes can hybridise with higher-order perpendicular standing spin wave (PSSW) modes, leading to anticrossing behaviour. This effect is particularly relevant for the $n = 0$ DE mode and the $n = 1$ PSSW mode in 200\,nm-thick YIG films~\cite{Riedel2023,Vilsmeier2024}. In the hybridisation region, the dispersion relation flattens and the group velocity approaches zero, resulting in suppressed SW propagation~\cite{Riedel2023,Vilsmeier2024}. Fig.~\ref{fig:SupplementaryMaterial_Figure5} shows a field–frequency map of the hybridisation-induced stop band, calculated using the micromagnetic simulation package \textsc{TetraX}~\cite{TetraX2022}, indicating the parameter range where SW propagation is inhibited. The stop band spans a broad excitation frequency range of $f\approx4.0$\,GHz to $f\approx4.25$\,GHz for external magnetic fields between $\mu_0H=68$\,mT and $\mu_0H=74$\,mT—matching the field range relevant from effective field considerations (cf. Fig.~\ref{fig:SupplementaryMaterial_Figure3}).

Analogous to Ref.~\cite{Vilsmeier2024}, local hybridisation effects can occur due to the inhomogeneous effective field distribution. At $f = 4.08$\,GHz (Fig.~\ref{fig:SupplementaryMaterial_Figure4}(c) and lower dotted line in Fig.~\ref{fig:SupplementaryMaterial_Figure5}), the stop band extends from $\mu_0H = 68$\,mT to $70.6$\,mT, matching the simulated effective field in the stripe, which is estimated to be around $\mu_0H_{x,\mathrm{eff}} \approx 71$\,mT or lower. As a result, SWs are strongly suppressed before even reaching the ring.

In Fig.~\ref{fig:SupplementaryMaterial_Figure4}(d), the DE mode propagates along the stripe but terminates near the ring entrance. At $f = 4.24$\,GHz (upper dotted line in Fig.~\ref{fig:SupplementaryMaterial_Figure5}), the lower field edge of the stop band is around $\mu_0H = 73.1$\,mT, which closely matches the iso-field contour in Fig.~\ref{fig:SupplementaryMaterial_Figure3}. This correspondence explains the observed stop position. However, since some SWs may scatter before reaching the stop band and the effective field does not drop below the threshold everywhere, weak signals may still reach the junction and beyond.

Overall, in the DE geometry, transmission is already limited by hybridisation-induced attenuation before interaction with the ring. As Fig.~\ref{fig:SupplementaryMaterial_Figure5} shows, this occurs over a broad frequency range, rendering the DE configuration less suitable for broadband SW filtering in 200\,nm thick YIG films.

\begin{figure}[h!]
    \includegraphics[width=0.7\textwidth]{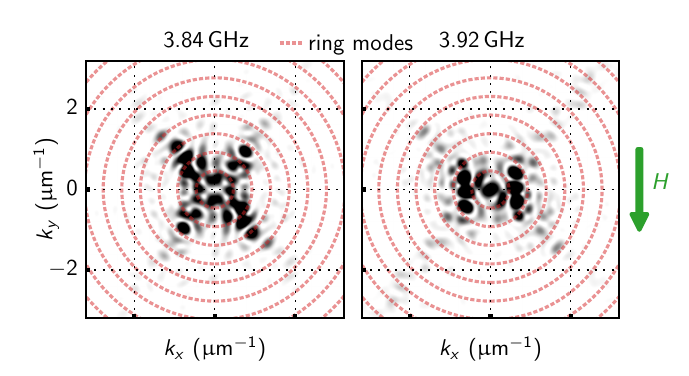}
    \caption{FT data extracted from ring section compared to ring eigenmodes. The FT data was calculated within the dashed green-colored contour in Figs.~\ref{fig:SupplementaryMaterial_Figure4}(a) and (b). Red dashed circles illustrate the ring eigenmodes across the width. For better contrast, the FT data is displayed on a logarithmic scale.}
    \label{fig:SupplementaryMaterial_Figure6}
\end{figure}

Fig.~\ref{fig:SupplementaryMaterial_Figure6} shows FT data extracted from the green contour in Figs.~\ref{fig:SupplementaryMaterial_Figure4}(a) and (b), corresponding to excitation frequencies of $f = 3.84$\,GHz and $f = 3.92$\,GHz, respectively. Only these two cases were considered, as they showed a sizable SW amplitude within the ring. Similar to the BV geometry, multiple SW modes appear, governed by the anisotropic dispersion relation and the inhomogeneous internal magnetic field, although the overall SW amplitude is less prominent. Due to the stronger spatial variation of the effective field in the DE configuration, the effective width of the ring is reduced. To account for this, a reduced width of $w_\mathrm{ring, eff} = 0.9\cdot w$ was used for the ring eigenmode calculation. The resulting FT spectra exhibit discrete features that align reasonably well with the adjusted eigenmodes. Overall, the observations suggest that transmission suppression in this frequency range results from scattering and interference within the ring section—consistent with the behaviour observed in the BV geometry, although the stronger demagnetising field in the DE case leads to a more complex mode structure.

\begin{figure}[h!]
    \includegraphics[width=\textwidth]{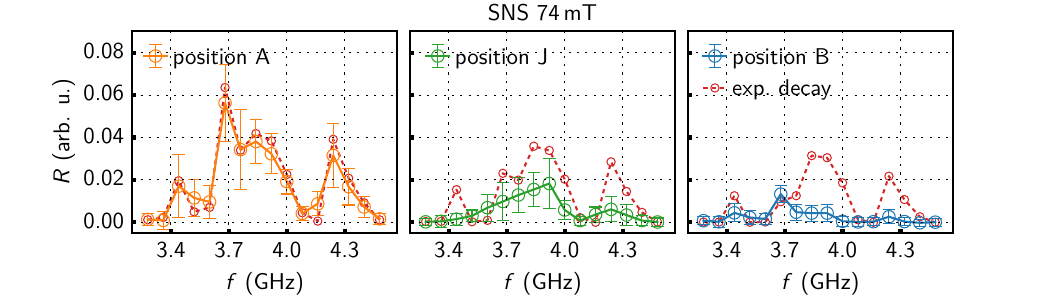}
    \caption{Spin wave amplitude averaged across regions A, J, and B as a function of excitation frequency in the DE geometry. The dip in amplitude at 4.08\,GHz is attributed to hybridisation-induced attenuation.}
    \label{fig:SupplementaryMaterial_Figure7}
\end{figure}

Further SNS-MOKE data were acquired at different excitation frequencies to evaluate the transmission properties, analogous to the BV case. Fig.~\ref{fig:SupplementaryMaterial_Figure7} presents the resulting amplitude signals averaged across regions A, J, and B in the DE geometry.

The amplitudes vary as a function of excitation frequency. In region A, a small peak is observed at $f = 3.44$\,GHz, with the maximum occurring at $f = 3.68$\,GHz. As the excitation frequency increases further, the amplitude in region A decreases, showing a pronounced dip at $f = 4.08$\,GHz, followed by a local maximum at $f = 4.24$\,GHz, and then a gradual decline. The amplitudes interpolated from an exponential fit in the stripe section (cf. Fig.~3(a) in the main text) generally match the measured values. However, this approach is less reliable in the DE geometry due to pronounced spatial variations in the internal effective field, particularly near the ring, which affect the local group velocity and thus the attenuation length.

The amplitude dip at $f \approx 4.08$\,GHz at position A can be directly attributed to the hybridisation-induced stop band, where increased damping prevents the DE mode from efficiently reaching region A (cf. Fig.\ref{fig:SupplementaryMaterial_Figure4}(c)). Similar arguments apply to the range between $f = 4.00$\,GHz and $f = 4.24$\,GHz for regions J and B, where coupling to the first PSSW mode significantly suppresses propagation, as discussed earlier.

In region J, the amplitude increases and peaks at $f = 3.92$\,GHz. A clear dip is observed at $f = 4.08$\,GHz, followed by a small local maximum at $f = 4.24$\,GHz. Notably, the extrapolated amplitude based on exponential decay deviates considerably at $f = 4.24$\,GHz.

In region B, a small peak appears at $f = 3.44$\,GHz and a maximum at $f = 3.68$\,GHz, both agree with the exponential fit. Beyond this frequency, however, the measured amplitude signal nearly vanishes, showing a strong discrepancy from the expected exponential behaviour. DE modes generally exhibit longer attenuation lengths than BV modes due to their higher group velocities, making the rapid decay observed here unexpected. While the suppression between $f = 4.00$\,GHz and $f = 4.24$\,GHz can be attributed to hybridisation-induced damping, the lack of transmission in the range from $f = 3.72$\,GHz to $f = 4.00$\,GHz cannot be explained by either hybridisation or intrinsic damping alone.

To conclude, SW propagation in the DE geometry is less frequency-selective and generally more strongly attenuated across the structure compared to the BV case. In 200\,nm thick YIG films, hybridisation with PSSW modes further limits transmission. Moreover, the behaviour is more complex due to pronounced demagnetisation effects. Nevertheless, caustic-like SW beams can still enable transmission across the ring.

%\newpage
%\printbibliography
\bibliography{supplement_bibliography}
\bibliographystyle{ieeetr}